\documentclass[english,12pt,a4paper]{article}
\usepackage{babel,indentfirst}
\usepackage{amssymb}

\begin{document}

\title{\bf Embeddings into the Pancake \\
Interconnection Network}
\author{Christian LAVAULT}

\date{\small \sl LIPN, CNRS ESA 7030, Universit\'e Paris~13 \\
99, av. J.-B. Cl\'ement 93430 Villetaneuse, France \\
{\rm E-mail: lavault@lipn.univ-paris13.fr}}
\maketitle

\vspace*{-.3cm}
\begin{abstract}
Owing to its nice properties, the pancake is one of the 
Cayley graphs that were proposed as alternatives to the 
hypercube for interconnecting processors in parallel 
computers. In this paper, we present embeddings of rings, 
grids and hypercubes into the pancake with constant 
dilation and congestion. We also extend the results 
to similar efficient embeddings into the star graph.
\end{abstract}

\newtheorem{thm}{Theorem} 
\newtheorem{defi}{Definition}
\newtheorem{lem}{Lemma}
\newtheorem{cor}{Corollary}
\newtheorem{cl}{Claim}
\newtheorem{prop}{Proposition} 
\newenvironment{proof}{\begin{trivlist} 
                       \item[]\hspace{0cm}{\bf Proof.} 
                       \hspace{0cm} }{\hfill $\square$ 
                       \end{trivlist}}
\newenvironment{ex}{%
		\begin{trivlist}{\setlength{\itemsep}{0cm}
		\item[]{\bf Example.}}%
		}{%
		\end{trivlist}}

\def\ra{\rightarrow}
\def\udl{\underline}
\def\b{\bar}
\newcommand{\bm}[1]{\mbox{\boldmath $ #1 $ }}

\bibliographystyle{article}
\def\bibfmta#1#2#3#4{{\bf #1}. {#2}, {\sl #3}, #4.}
\bibliographystyle{book}
\def\bibfmtb#1#2#3#4{{\bf #1}. {\sl #2}, { #3}, #4.}

\section{Introduction}
Akers and Krishnamurthy~\cite{AK89} proposed the pancake 
and the star as alternatives to the hypercube for interconnecting 
processors in parallel computers. These networks have some nice 
properties: edge and vertex symmetry (strong symmetry), small 
degree and diameter, extensibility, high connectivity (robustness), 
easy routings and broadcasting, etc. To compare favorably with 
the hypercube, these graphs must also offer good and simple 
simulations of other interconnection networks. The problem of 
simulating known networks by the star graph has been extensively 
studied. For example, Nigam, Sahni and Krishnamurthy~\cite{NSK90} 
consider embeddings of rings and hypercubes in star graphs. 
Miller, Pritkin, and Sudborough~\cite{MPS94} study one-to-one 
and one-to-many embeddings of hypercubes into Cayley graphs. 
Jwo, Lakshmivarahan, and Dhall~\cite{JLD90}, Qiu, Meijer, 
and Akl~\cite{QMA93,QAM94} consider embeddings of grids in stars. 
Bouabdallah, Heydemann, Opatrny and Sotteau~\cite{BHOS94} 
present embeddings of complete binary tree into star networks. 
Azevedo, Bagherzadeh and Latifi~\cite{ABL96} propose embeddings 
of hypercubes in star graphs.

However, embedding interconnection networks into the pancake 
has received less attention. In this paper, we focus on the 
problem and present embeddings of rings, grids and hypercubes 
into the pancake with constant dilation and congestion.

\medskip
The paper is organized as follows. In the Preliminaries, 
we state the definitions and the group-theoretic terminology 
that are used in this paper. Section~3 is devoted to embeddings 
of rings and arrays into the pancake. In Section~4, we present 
embeddings of grids into the pancake and the star, and in Section~5, 
we consider embeddings of hypercubes (binary and generalized 
hypercubes) into the pancake; the results are extended to the star. 
The concluding Section~6 briefly outlines possible improvements 
and open problems.

\section{Preliminaries}
Following~\cite{AK89,B74}, we first present the group-theoretic 
model used to design and analyze the pancake. Next we define 
the pancake network itself.

\begin{defi}
Let $\cal G$ be a finite multiplicative group. 
Let I be the identity in $\cal G$ and G a set of 
generators of $\cal G$ with the following two properties

(i)\ $(\forall g\in G) \quad g^{-1}\in G$;

(ii)\ $I\notin G$.

Given $({\cal G},G)$, a Cayley graph $(V,E)$ is defined 
as a simple graph, whose vertex-set and edge-set~are 
$$V={\cal G}\ \quad \mbox{and}\ \quad 
E=\{(u,v)\in V\times V\;|\;u^{-1}v\in G\}.$$
\end{defi}

It is easily seen that Cayley graphs $(V,E)$ are finite, 
connected, undirected, devoid of multiple edges, loop-free, 
and symmetric. Since interconnection networks may be viewed 
as an undirected graph, we will use the terms graph and 
(interconnection) network interchangeably.

\bigskip \noindent
{\bf Notation.}\, In the remainder of the paper, we use 
the usual terminology of basic group theory and graph theory. 
Since we only consider finite groups, the groups are mainly 
represented as {\em permutation groups}. 
The following notation is used:
\begin{itemize}
\item Let $X$ and $Y$ be two sets, $X\setminus Y$ denotes 
the relative complement of the set $Y$ with respect to the 
set $X$.
\item $S_n$ is the {\em symmetric group} on $n$ symbols, 
i.e. on $\{1,\ldots,n\}$ for simplicity. The multiplication 
in $S_n$ is the composition of permutations. 

A permutation $\pi\in S_n$ is denoted by $x_1 x_2\cdots x_n$, 
where we have ${\pi}(k)=x_k$ for $k=1,\ldots,n$. 
This representation is referred to as the {\em standard 
representation} of a permutation, to distinguish it from 
other representations that are introduced further. 
The identity permutation on $n$ symbols is thus $I=123\cdots n$.
\item $\tau$ denotes the {\em transposition permutation} 
and ${\sigma}(\ell,m)$ denotes the {\em cyclic permutation} 
of $m$ positions over the first $\ell$ symbols: 
$${\sigma}(\ell,m)=(\ell -m+1)\cdots \ell 1\cdots 
(\ell -m)(\ell +1)\cdots n.$$
\item Symbols are denoted by lower case letters, and 
{\em blocks} of symbols by upper case letters. No ambiguity 
may rise from the notation $I$, since $I$ is (by definition) 
the unique block of symbols $123\cdots n$.
\item A permutation $\pi=x_1 x_2\cdots x_{i-1} x_ix_{i+1}\cdots x_n$ 
can be represented by blocks of symbols instead of its 
symbols themselves. For example, we can write $\pi=A\,B$, 
where $A=x_1 x_2\cdots x_{i-1} x_i$ and $B=x_{i+1}\cdots x_n$.

For any block of symbols $A$, $\b{A}$ denotes the block 
obtained by reversing $A$. For example, if 
$\rho=x_i x_{i-1}\cdots x_2 x_1x_{i+1}\cdots x_n$, 
we write $\rho=\b{A}\,B$, where $A$ and $B$ are the 
blocks defined above.
\end{itemize}

\begin{defi} \,{\em \cite{AK89}} \\ \label{defpan}
The pancake network $P_n$ of dimension $n$ is the Cayley 
graph $(S_n,E)$, whose set of generators is 
$$G=\{g_i\in S_n\;|\;g_i=i(i-1)\cdots 321(i+1)\cdots n,\;i=2,\ldots,n\}.$$

In other words, the $n!$ vertices of $P_n$ are labeled with 
the permutations on n symbols (of $S_n$), and any two vertices 
of $P_n$, $u=x_1 x_2\cdots x_n$ and $v~=~y_1~y_2~\cdots~y_n$, 
are connected (i.e. $(u,v)\in E$) iff there exists an integer i, 
$2~\le~i~\le~n$, such that $y_j=x_{i-j+1}$ for $j=1,\ldots,i$, 
and $y_j=x_j$ for $j>i$.
\end{defi}

Clearly, there are $(n-1)$ generators, one for each value of 
$i$, $2\le i\le n$, and $|G|=n-1$. It is easy to show that 
the Cayley network $P_n$ has $n!$ vertices, each with degree 
$|G|=n-1$: $P_n$ is $(n-1)$-regular.

When a permutation $\rho$ is obtained from a permutation 
$\pi=x_1\cdots x_n$ by applying a generator $g_i\in G$, 
we write $x_1\cdots x_i\cdots x_n$\, $\ra$\, 
$\udl{x_i\cdots x_1}\cdots x_n$.

\medskip
$P_n$ can be decomposed into $n$ subpancakes each of dimension 
$(n-1)$. Each of the $(n-1)!$ vertices of each subpancake has 
a block representation of the form $Ai$, where $A\in S_{n-1}$ 
is a ``permutations block'' on the $(n-1)$ symbols 
$\{1,\ldots,n\}\setminus \{i\}$, for a given 
$i\in \{1,\ldots,n\}$ which depends on the considered 
subpancake. As a consequence, each of the $n$ subpancakes 
of $P_n$ (one for each value of $i$, $1\le i\le n$) 
can be represented with two distinct notations depending 
on the context:
\begin{enumerate}
\item For a given integer $i$ $(1\le i\le n)$, $P_{n-1,i}$ 
denotes the subpancake defined from the above representation, 
i.e. $i$ is the last symbol of each vertex/permutation 
of $P_{n-1,i}$.
\item  For a given permutation $\pi\in P_{n-1,i}$ 
$(1\le i\le n)$, $P_{n-1}(\pi)$ denotes the subpancake 
defined from the element $\pi$, i.e. $\pi$ is the 
representation of a vertex of that subpancake.
\end{enumerate}
A subpancake of dimension $k$ is called a $k$-pancake.

\begin{defi} \label{defstar}
The star network of dimension $n$ (or n-star) is the Cayley 
graph whose set of generators is $G=\{\tau(1,i)\;|\;i=1,\ldots,n\}$, 
where $\tau\in S_n$ is the transposition permutation. 
In other words, the $n!$ vertices of the n-star are labeled 
with the permutations on n symbols, and each vertex $u$ 
of the n-star, $u=x_1 x_2\cdots x_n$ is connected to the 
$n-1$ vertices v such that 
$v=x_i x_2\cdots x_{i-1}x_1 x_{i+1}\cdots x_n$, 
for $i=2,\ldots,n$.
\end{defi}
Now recall the definition of an embedding of a graph 
into another graph.
\begin{defi}
Given two undirected connected graphs $H_1=(V_1,E_1)$ and 
$H_2=(V_2,E_2)$ such that $|V_1|\le |V_2|$, the embedding 
of $H_1$ into $H_2$ maps $V_1$ into $V_2$. The ratio 
$|V_2|/|V_1|$ is the expansion of the embedding. 
The dilation of any edge $(x_1,y_1)\in E_1$, is the length 
of the path $[x_2,y_2]$, where $x_1\mapsto x_2$ and 
$y_1\mapsto y_2$ in the embedding, respectively. 
The dilation of the embedding is the maximum over all dilations. 
The congestion of an edge $(x_2,y_2)\in E_2$ is the number 
of edges $(x_1,y_1)\in E_1$ whose image by the mapping contains 
$(x_2,y_2)$. The congestion of the embedding is the maximum 
over all congestions.
\end{defi}

\section{Embeddings of Rings}
In this section we consider rings of size $k!$ ($3\le k\le n$). 
The dilation and the congestion of embeddings of such rings into 
$P_n$ are shown to be 1.
\begin{prop}
For any integer i, $2\le i\le n-1$, and any generator $g_i\in G$ 
as defined in Definition~\ref{defpan}, 
$\underbrace{g_ig_{i+1}\cdots g_ig_{i+1}}_{i+1}=I$.
\end{prop}
\begin{proof}
Let a permutation 
$\pi=x_1 x_2\cdots x_{i-1} x_i x_{i+1}\cdots x_n=A\, x_{i+1}\, B$. 
Then $\rho=\pi\, g_i=\b{A}\, x_{i+1}\,B$, and 
$$\pi\, g_i g_{i+1}\;=\;\rho\, g_{i+1}=x_{i+1}\, A\,B\;=\; 
x_{i+1} x_1 x_2\cdots x_{i-1} x_i x_{i+2}\cdots x_n.$$
In other words, $g_i g_{i+1}$ is the cyclic permutation 
${\sigma}(i+1,1)$, and since ${\sigma}(i+1,1)^{i+1}=I$, 
the result follows.
\end{proof}

For any permutation $\pi$ and any sequence of generators 
$H=h_1,\ldots,h_k$, we denote by $(\pi,H)$ the corresponding sequence 
of permutations ${\pi}_0,\ldots,{\pi}_k$ such that ${\pi}_0=\pi$ and 
$\pi_i=\pi_{i-1}\,h_i$, for all $i=1,2,\ldots,k$.

\begin{defi}
For $k=2,\ldots,n$, the pancake sequence $G_k$ of order k is the 
sequence of generators recursively defined as follows:

(i)\ $G_2=g_2$;

(ii)\ for $k>2$, 
$G_k=\langle G_{k-1},g_k,G_{k-1},g_k,\ldots,G_{k-1}\rangle$, 
where $G_{k-1}$ occurs k times in the sequence.
\end{defi}

\begin{prop} \label{hami}
Given a permutation $\pi\in S_n$, for $k=3,\ldots,n$, $(\pi,G_k)$ 
defines a Hamiltonian cycle over the k-pancake containing 
$\pi$. In this Hamiltonian cycle, the vertices of each 
subpancake have adjacent locations.
\end{prop}
\begin{proof}
The proof is by induction on $k$. Since the pancake is 
vertex transitive, we assume that $\pi=I$.

\medskip\noindent
$\bullet$ {\bf Base $(k=3)$:}\ Applying the generators 
of $G_3$ yields the following sequence of permutations: 

$I=123\cdots n\,\ra\, \udl{21}3\cdots n\, 
\ra\, \udl{312}\cdots n\, \ra\, \udl{13}2\cdots n\, 
\ra\, \udl{231}\cdots n\, \ra\, \udl{32}1\cdots n$.

\smallskip \noindent
It is easily verified that all the elements of $P_3(I)$ 
belong to the sequence and that the last element of the 
list is connected to the first one through the generator 
$g_3$.

\medskip\noindent
$\bullet$ {\bf Induction step:}\ Suppose that $(\pi,G_{n-1})$ 
defines a Hamiltonian cycle over $P_{n-1}(\pi)$.
We first show that the permutation obtained by applying 
$\langle G_{n-1},g_n,\ldots,G_{n-1},g_n\rangle$ ($h$ times) 
is $(n-h+1)(n-h+2)\cdots n12\cdots (n-h)$. The property 
holds for $h=1$ since the permutation is obtained from 
the sequence 

\medskip
$\qquad \qquad  12\cdots (n-1)n\, \ra\, \udl{(n-1)\cdots 21} n\, 
\ra\, \udl{n12\cdots (n-1)}$.

\medskip\noindent
Let us now suppose that the property holds up to $h$. 
The next step is then

\medskip
$\qquad \ (n-h+1)(n-h+2)\cdots n12\cdots (n-h-1)(n-h)\,\ra$
 
$\udl{(n-h-1)\cdots 21n\cdots (n-h+1)}(n-h)\;\ra 
\;\udl{(n-h)\cdots n12\cdots (n-h-1)}$.

\medskip \noindent
Therefore, after applying $\langle G_{n-1},g_n,\ldots,G_{n-1},g_n\rangle$ 
($h$ times), the permutation corresponds to a vertex of $P_{n-1}(n-h)$.
According to the induction step, the next $G_{n-1}$ visits all
the vertices of the subpancake $P_{n-1}(n-h)$. Whence the result 
that all vertices of all the $(n-1)$-pancakes in $P_n$ are visited. 
The last visited vertex is $(2\cdots n1)\,g_{n-1}= n\cdots 21$, 
and this permutation is connected to $I$ through $g_n$: 
the proof is completed.
\end{proof}

In the following, we still let $\pi=I$ (w.l.o.g.). The order 
relation induced by the sequence $(I,G_n)$ on permutations will 
be referred to as the ordering of the pancake sequence.
\begin{ex}
Let $n= 4$. The list of vertices of $P_4$ ordered by the pancake 
sequence is: \\
$1234$ $\ra$ $\udl{21}34$ $\ra$ $\udl{312}4$ 
$\ra$ $\udl{13}24$ $\ra$ $\udl{231}4$ $\ra$ $\udl{32}14$ 
$\ra$ $\udl{4123}$ $\ra$ $\udl{14}23$ $\ra$ $\udl{241}3$ 
$\ra$ $\udl{42}13$ $\ra$ $\udl{124}3$ $\ra$ $\udl{21}43$ 
$\ra$ $\udl{3412}$ $\ra$ $\udl{43}12$ $\ra$ $\udl{134}2$ 
$\ra$ $\udl{31}42$ $\ra$ $\udl{413}2$ $\ra$ $\udl{14}32$ 
$\ra$ $\udl{2341}$ $\ra$ $\udl{32}41$ $\ra$ $\udl{423}1$ 
$\ra$ $\udl{24}31$ $\ra$ $\udl{342}1$ $\ra$ $\udl{43}21$.
\end{ex}
Theorem~\ref{ringk} easily follows.

\begin{thm} \label{ringk}
For $k=3,\ldots,n$, the ring of size $k!$ can be embedded into 
the $n$-pancake with dilation $1$ and congestion $1$.
\end{thm}
\begin{proof}
Immediate from Proposition~\ref{hami}. Given a Hamiltonian 
graph of order $n$, the corresponding ring can be embedded 
into that graph with dilation and congestion~1.
\end{proof}
As a consequence of Theorem~\ref{ringk}, we also have the 
\begin{cor}
For $\ell$ such that $\ell\le n!$, the linear array (line) 
of length $\ell$ can be embedded into the $n$-pancake with 
dilation $1$ and congestion $1$.
\end{cor}

\section{Embeddings of Grids}
\subsection{Embeddings of $\bm{N_1\times N_2}$ Grids}
Given any two positive integers $N_1$ and $N_2$, we first 
consider embeddings of $N_1\times N_2$ grids with 
$N_1 N_2\le n!$ into $P_n$ and give a negative result.

\begin{prop}\label{subgrid}
The $N_1\times N_2$ grid is not a subgraph of the n-pancake.
\end{prop}
\begin{proof}
The proof is by contradiction. The $2\times 2$ grid is a 
subgraph of the $N_1\times N_2$ grid. Suppose the $2\times 2$ 
grid were a subgraph of $P_n$, then there would be two 
permutations $X$ and $Y$, and four generators $g_i$, $g_j$, 
$g_\ell$, $g_k$, with $i\ne j$, $i\ne k$, $k\ne \ell$, 
such that $Y=X\,g_i$ and $Y\,g_k=X\,g_j\,g_\ell$. 
Hence, $g_i\,g_k=g_j\,g_\ell$, which would imply that 
$j=\ell$ and $k=i$, or $j=i$ and $\ell=k$: a contradiction.
\end{proof}

\begin{lem} \label{lm1}
For any two integers $\ell$ and $m$ such that 
$0\le m\le \ell\le n$, the cyclic permutation $\sigma(\ell,m)$ 
can always be built with two or three generators of the pancake.
\end{lem}
\begin{proof}
Let a permutation 
$\pi=x_1\cdots x_{\ell-m} x_{\ell-m+1}\cdots x_\ell x_{\ell+1}\cdots 
x_n=ABC$, with blocks 
$A=x_1\cdots x_{\ell-m}$, $B=x_{\ell-m+1}\cdots x_\ell$\ and 
$C=x_{\ell+1}\cdots x_n$. Then,

\noindent
$\pi\,\sigma(\ell,m)=x_{\ell-m+1}\cdots~x_\ell~x_1\cdots 
x_{\ell-m}~x_{\ell+1}\cdots~x_n=BAC$,\ and we have the following 
path joining $\pi$ to $\pi\,\sigma(\ell,m)$:\ 
$\pi=ABC\ra \b{A}BC \ra \b{B}AC\ra BAC=\pi\,\sigma(\ell,m)$.
The length of this path is 3 when $1<m<\ell-1$, and it is 2 
whenever $m=1$ or $m=\ell-1$.
\end{proof}

Now from Lemma~\ref{lm1} we present an embedding of the 
$n\times (n-1)!$ grid in $P_n$ with constant dilation.

\begin{thm}
The $n\times (n-1)!$ grid can be embedded in the n-pancake 
with dilation~$7$.
\end{thm}
\begin{proof}
The first row of the grid is represented by the first 
$(n-1)$-pancake ordered from the pancake sequence. 
For $0\le j\le (n-1)!\,-\,1$, let ${\pi}_j$ be the vertex 
of the pancake corresponding to the node $(0,j)$ on the grid. 
A node $(i,j)$, with $i\ne 0$, is represented by 
$\pi_j\,\sigma(n,i)$. Now, considering two adjacent nodes 
on the grid, let us compute the distance between those 
vertices of the pancake that represent them.
\begin{itemize}
\def\labelitemi{$\bullet$}
\item Two nodes $(0,j)$ and $(0,j+1)$ are represented 
by two adjacent vertices of the pancake.
\item Two nodes $(i,j)$ and $(i+1,j)$ are represented 
by the two vertices $X=\pi_j\,\sigma(n,i)$ and 
$Y=\pi_j\,\sigma(n,i+1)=X\,\sigma(n,1)$. 
According to Lemma~\ref{lm1}, the distance between $X$ 
and $Y$ is 2.
\item Two nodes $(i,j)$ and $(i,j+1)$ are represented 
by the two vertices $Y_1=\pi_j\,\sigma(n,i)$ and 
$Y_2=\pi_{j+1}\,\sigma(n,i)$. Then, for a generator $g_k$, 
$\pi_{j+1}=\pi_j\,g_k$. To compute the distance between $Y_1$ 
and $Y_2$, two distinct cases (and two subcases) may arise:
\end{itemize}
\begin{description}
\item[-- First case:]\ $\pi_j=ABC$, $\pi_{j+1}=\b{A}BC$ 
and $Y_1=CAB$. Then, $Y_2=C\b{A}B$, and a path joining 
$Y_1$ to $Y_2$ is $CAB$ $\ra$ $\b{A}\b{C}B$ $\ra$ $A\b{C}B$ 
$\ra$ $C\b{A}B$. The distance from $Y_1$ to $Y_2$ is 3.
\item[-- Second case:]\ $\pi_j=ABC$, $\pi_{j+1}=\b{B}\b{A}C$, 
and $Y_1=BCA$. To obtain $Y_2$, two subcases must be 
considered.
\begin{description}
\item[- First subcase:]\ $Y_2=\b{B_1} \b{A} C \b{B_2}$. 
In this subcase, a path from $Y_1$ to $Y_2$ is 
$Y_1=B_1 B_2 C A$\, $\ra$\, $\b{B_2} \b{B_1} C A$\, 
$\ra$\, $B_2 \b{B_1} C A$\, $\ra$\, 
$\b{A} \b{C} B_1 \b{B_2}$\, $\ra$\, 
$C A B_1 \b{B_2}$\, $\ra$\, $\b{C} A B_1 \b{B_2}$\, 
$\ra$\, $\b{B_1} \b{A} C \b{B_2} = Y_2$. 
The distance from $Y_1$ to $Y_2$ is 6.
\item[- Second subcase:]\ $Y_2=\b{A_1} C \b{B} \b{A_2}$. 
In this last subcase a path from $Y_1$ to $Y_2$ is 
$Y_1=B C A_1 A_2$\, $\ra$\, $\b{A_2} \b{A_1} \b{C} \b{B}$\, 
$\ra$\, $A_2 \b{A_1} \b{C} \b{B}$\, 
$\ra$\, $B C A_1 \b{A_2}$\, 
$\ra$\, $\b{C} \b{B} A_1 \b{A_2}$\, 
$\ra$\, $C \b{B} A_1 \b{A_2}$\, 
$\ra$\, $B \b{C} A_1 \b{A_2}$\, 
$\ra$\, $\b{A_1} C \b{B} \b{A_2}=Y_2$.
\end{description}
\end{description}
This last configuration yields a (maximal) distance 7 from 
$Y_1$ to $Y_2$.
\end{proof}

The same method applies to the $(n+(n-2)+(n-3)+\cdots+(p+1))\times p!$  
grid, for $p=2,\ldots,n-1$; the following Theorem~\ref{grid1} 
shows that this grid can be embedded into $P_n$ with constant 
dilation. Note that the term $(n-1)$ is omitted in the definition 
of the grid. Indeed, by Proposition~\ref{subgrid}, we already know 
that the $(n+(n-1)+(n-2)+\cdots+(p+1))\times p!$ grid 
($2\le p\le n-1$) is not a subgraph of $P_n$. 

\begin{thm} \label{grid1}
For $p=2,\ldots,n-1$, the $(n+(n-2)+(n-3)+\cdots+(p+1))\times p!$ 
grid can be embedded in the n-pancake with dilation $4$.
\end{thm}
\begin{proof}The first row of the mesh is represented by 
the first $p$-pancake. The next $(n-1)$ rows are obtained 
by applying the cyclic permutations $\sigma(n,i)$ to the 
first row. The next $(n-2)$ rows are obtained by applying 
the cyclic permutations $\sigma(n-1,i)$ to the first row, etc. 
Finally, the last $p$ rows are obtained by applying the cyclic 
permutations $\sigma(p+1,i)$ to the first row. 
The only new adjacent nodes to consider are $X\,\sigma(k,k-1)$ 
and $X\,\sigma(k-1,1)$. Let $X=x_1\,A\,x_{k-1}\,x_k\,B$. 
Then, $X\,\sigma(k,k-1)=A\,x_{k-1}\,x_k\,x_1\,B$ 
and $X\,\sigma(k-1,1)=x_{k-1}\,x_1\,A\,x_k\,B$. 
A path joining these two vertices is
\begin{eqnarray*}
\lefteqn{A\,x_{k-1}\,x_k\,x_1\,B\; \ra\; x_k\,x_{k-1}\,\b{A}\,x_1\,B\; 
\ra\; x_1\,A\,x_{k-1}\,x_k\,B} \\ 
& & \ra\; \b{A}\,x_1\,x_{k-1}\,x_k\,B\; \ra\; x_{k-1}\,x_1\,A\,x_k\,B.
\end{eqnarray*}
The distance between the two vertices is thus 4 
and the proof follows.
\end{proof}

\subsection{Embeddings of $\bm{n}$-Grids}
For embedding $n$-grids into $P_n$, a new representation 
of permutations is first introduced. A permutation $\pi$ 
may be represented as $\pi=a_2a_3\cdots a_n$, where $a_i$ 
is the number of symbols less than $i$ that are located 
at the left of $i$ in the standard representation of $\pi$.

\begin{ex}
Let $n=5$. The permutation 12345 is represented by 1234, 
and the permutation 54321 is represented by 0000. 
Similarly, the permutation 42153 is represented by 0203.
\end{ex}

The map 
$S_n\,\longrightarrow\,\{\pi=a_2,\ldots,a_n\;|\; 0\le a_i\le i-1\}$ 
is obviously one-one, and it is used in this subsection 
to embed the $2{\times}3{\times}\cdots {\times}(n-1){\times}n$ 
grid into $P_n$.

\begin{thm}
The $2{\times}3{\times}\cdots {\times}(n-1){\times}n$ grid 
can be embedded into the n-pancake with dilation $6$.
\end{thm}
\begin{proof} 
Let two vertices $X$ and $Y$ on the grid be labeled with the 
two permutations $a_2\cdots a_{i-1}{\alpha}a_{i+1}\cdots a_n$ 
and $a_2\cdots a_{i-1}{\beta}a_{i+1}\cdots a_n$, respectively. 
$X$ and $Y$ are connected on the grid iff $\alpha=\beta+1$ 
or $\alpha=\beta-1$. Let us find the distance from $X$ to $Y$ 
in the pancake.

\noindent
We may assume w.l.o.g. that $\alpha=\beta+1$. 
Let $X=A\,x_k\,B\,i\,C$, where $x_k$ is the $\alpha$th 
symbol $<i$ and all symbols in $B$ are $>i$. 
Consider the permutation $A\,i\,B\,x_k\,C=b_2\cdots b_n$ 
and compare the $a_j\/$s and the $b_j\/$s.

\noindent
First, for all symbols in $A$ and $C$, $b_j=a_j$. Next, 
for each symbol $j$ in $B$, a new symbol that is smaller 
than $j$ is located on the left of $j$: it is the symbol $i$. 
Similarly, a new symbol that is smaller than $j$ is located 
on the right of $j$: it is the symbol $x_k$. 
Hence, $b_j=a_j+1-1=a_j$. New symbols located on the left 
of $x_k$ are larger than $x_k$; they are either the symbol 
$i$ or any symbol $j>i$, and hence, $b_{x_k}=a_{x_k}$. 
There is only one new symbol smaller than $i$ located 
on the right of $i$: it is the symbol $x_k$, other symbols 
are larger than $i$. Hence, $b_i=a_i-1=\alpha-1=\beta$.

\noindent
Therefore, $A\,i\,B\,x_k\,C=Y$, and a path joining 
$X$ to $Y$ is
\begin{eqnarray*}
\lefteqn{X\;=\;A\,x_k\,B\,i\,C\; \ra\; x_k\,\b{A}\,B\,i\,C\; 
\ra\; i\,\b{B}\,A\,x_k\,C\; \ra\; B\,i\,A\,x_k\,C} \\ 
& & \ra \; \b{B}\,i\,A\,x_k\,C\; \ra\; \b{A}\,i\,B\,x_k\,C\; 
\ra\; A\,i\,B\,x_k\,C\;=\; Y.
\end{eqnarray*}
The distance from $X$ to $Y$ is thus 6, and the dilation 
follows.
\end{proof}
The following corollary is easily derived.

\begin{cor}
The binary hypercube $Q_n$ can be embedded in the n-pancake 
with dilation~$6$.
\end{cor}
\begin{proof}
$Q_n$ is a subgraph of the $2\times 3\times\cdots \times (n-1)\times n$ 
grid.
\end{proof}

The same method applies to embed the $n$-grid into 
the star graph (see Definition~\ref{defstar}).

\begin{thm}
The $2\times 3\times \cdots \times (n-1)\times n$ grid 
can be embedded into the n-star with dilation $3$.
\end{thm}
\begin{proof}
The above representation of permutations is used again. 
Consider two permutations $X=aAxByC$ and $Y=aAyBxC$, and 
compute the distance from $X$ to $Y$ within the $n$-star. 
A path joining $X$ to $Y$ is
$$X=aAxByC\, \ra\, yAxBaC\, \ra\, xAyBaC\, \ra\, aAyBxC=Y.$$ 
The distance from $X$ to $Y$ is thus 3, and the dilation 
follows.
\end{proof}

\section{Embeddings of the Generalized Hypercube}
The $2\times 3\times \cdots \times (n-1)\times n$ 
{\em generalized hypercube (GHC)}~\cite{BA84} is the graph 
$(V,E)$ whose vertices are labeled withthe permutations 
$x_2,\ldots,x_n$, where $0\le x_i\le i-1$. Any two vertices 
$u,\,v\in V$ are connected iff their labels differ in only 
one position: 
i.e. there is an edge $(u,v)\in E$ between the two vertices 
$u=x_2,\ldots,x_{i-1},\alpha,x_{i+1},\ldots,x_n$ and 
$v=x_2,\ldots,x_{i-1},\beta, x_{i+1},\ldots,x_n$ iff 
$\alpha\ne \beta$ for some symbols $\alpha$ and $\beta$ 
from $x_2$ onwards.

Embedding the GHC into the $n$-pancake could be performed 
via the representation of permutations defined in the 
previous section. Unfortunately, the resulting dilation 
is $O(n)$, i.e. the dilation would then have the same order 
of magnitude as the diameter of the pancake. Consequently, 
a more suited representation of permutations must be used.

\medskip
A permutation $\pi=x_1\cdots x_n$ is represented by 
$a_2\cdots a_n$ with the following rule~{\bf (R)}:

\smallskip \noindent
{\bf for}\ $k=n$\ {\bf to}\ 2\ (step $-1$)\ {\bf do}

$\quad$\ $a_k\gets x_k - 1$~; $x_k\leftrightarrow k$\ \ 
({\em i.e.~exchange symbols $x_k$ and $k$ in permutation} 
$\pi$).

\begin{ex}
Let $n=8$ and $X=27351864$.Applying rule~{\bf (R)} 
step by step yields

$\quad$\ $a_8=4-1=3$, $\quad$ and\ $Y_1=27351468$,

$\quad$\ $a_7=6-1=5$, $\quad$ and\ $Y_2=26351478$,

$\quad$\ $a_6=4-1=3$, $\quad$ and\ $Y_3=24351678$,

$\quad$\ $a_5=1-1=0$, $\quad$ and\ $Y_4=24315678$,

$\quad$\ $a_4=1-1=0$, $\quad$ and\ $Y_5=21345678$,

$\quad$\ $a_3=3-1=2$, $\quad$ and\ $Y_6=21345678$,

$\quad$\ $a_2=1-1=0$, $\quad$ and\ $Y_7=12345678$. 
The representation of $X$ is $0200353$.
\end{ex}

\begin{prop}
The representation given above defines a one-one mapping 
between the n-pancake and the 
$2\times 3\times\cdots\times (n-1)\times n$ generalized 
hypercube.
\end{prop}
\begin{proof}
Let $a_2,\ldots,a_p,p,\ldots,n-1$ ($2\le p\le n-1$) 
denote the above representation of a permutation. 
The proof is by induction on $p$.

\medskip
$\bullet$ {\bf Base:}\ Let $p=2$ and $X=x_1x_234\cdots (n-1)n$ 
be a permutation. In that case, $\{x_1,x_2\}= \{1,2\}$. 
If $x_1=1$, from rule {\bf (R)} we have $a_i=i-1$ for 
each value of $i$ (in particular, $a_2=1$). Similarly, 
if $x_1=2$ we have $a_2=0$. Hence, the property holds 
for $p=2$.

\medskip
$\bullet$ {\bf Induction step:}\ Given $p$\ $(2\le p\le n-1)$, 
suppose the map from the set of permutations 
$\{x_1\cdots x_p(p+1)\cdots n\}$ onto the subgraph of the GHC 
defined by $\{a_2,\ldots,a_p,p,\ldots,n-1\}$ is one-to-one.
Let us prove that the property also holds for $(p+1)$.

Consider the permutations 
$X=x_1\cdots x_px_{p+1}(p+2)\cdots n$, and notice that the 
symbol $(p+1)$ belongs to the set $\{x_1,\ldots,x_p x_{p+1}\}$. 
According to rule~{\bf (R)}, the representation of $X$ is 
constructed step by step, from $i=n$ downto $i=2$, by performing 
$a_i=x_i-1$ and exchanging symbols $x_i$ and $i$ within $X$. 
Therefore, $x_{p+1}$ and $(p+1)$ are exchanged in the 
representation of $X$ and $a_{p+1}$ takes all the values 
0, 1, \ldots, $p$.

\noindent
Since $X$ is of the form $x_1\cdots x_p(p+1)\cdots n$, 
and according to the induction step, the values $a_i$ 
$(1<i<p+1)$ cover the whole set $\{0,\ldots,i-1\}$. 
Whence the property holds for $(p+1)$.
\end{proof}

Now, let $X=a_2\cdots a_{i-1}{\alpha}a_{i+1}\cdots a_n$ 
and $Y=a_2\cdots a_{i-1}{\beta}a_{i+1}\cdots a_n$ 
be the above representations of two permutations. 
To find the distance from $X$ to $Y$ within the 
pancake $P_n$ we need Lemma~\ref{lm2} first.

\begin{lem} \label{lm2}
Let two permutations $X$ and $Y$ denoted 
$a_2\cdots a_{i-1}{\alpha}a_{i+1}\cdots a_n$ and 
$a_2\cdots a_{i-1}{\beta}a_{i+1}\cdots a_n$, 
respectively. Their standard representations differ 
in at most three positions, i.e. $X=AxByCzD$ and 
$Y=AzBxCyD$.
\end{lem}
\begin{proof}
Consider two permutations $W$ and $Z$, such that 
$W=AxBpC$, where the symbol $p$ is located at position 
$p$, with $p>x$, and $Z=ApBxC$. Let us compare each 
of the respective representations of $W$ and $Z$.

In each representation, the values of $a_i$ corresponding 
to $C$ are equal; the values of $a_p$ are $(p-1)$ and $x-1$, 
and for $i<p$, $a_i$ is obtained from rule~{\bf (R)}. 
The values of the $a_i\/$s within each representation of $W$ 
and $Z$ are equal to the symbols in two 
permutations $\pi_W$ and $\pi_Z$ (respectively), each obtained 
by applying rule~{\bf (R)}. Now, this construction of $\pi_W$ 
and $\pi_Z$ yields $\pi_W=\pi_Z=ExFp(p+1)\cdots n$, where $E$ 
and $F$ are two blocks of symbols in 
$\{1,\ldots,x\}\setminus \{x,p,\ldots,n\}$. 
Hence, the representations of $W$ and $Z$ differ in 
one position only and, for a given $X$, there are 
$(p-1)$ such $Y\/$s. The standard representations of 
two such $Y\/$s differ in three positions, and the proof 
follows.
\end{proof}
The following Theorem~\ref{hypan} derives from Lemma~\ref{lm2}.

\begin{thm} \label{hypan}
The $2\times 3\times \cdots \times (n-1)\times n$ 
generalized hypercube can be embedded into the n-pancake 
with dilation $8$.
\end{thm}
\begin{proof}
Let two permutations $X$ and $Y$, whose representations 
differ in one position only. According to Lemma~\ref{lm2}, 
each of their standard representations differs in at most 
three positions, i.e. $X=AxByCzD$\ and $Y=AzBxCyD$. 
A path joining $X$ to $Y$ is thus

$\qquad X=AxByCzD\, \ra\, z\b{C}y\b{B}x\b{A}D\, \ra\, xByCz\b{A}D\, 
\ra\, \b{B}xyCz\b{A}D\,\ra\,$ 

$BxyCz\b{A}D \, \ra\, \b{C}yx\b{B}z\b{A}D\, \ra\, Cyx\b{B}z\b{A}D\, 
\ra\, y\b{C}x\b{B}z\b{A}D\, \ra\, AzBxCyD=Y$,

\smallskip \noindent
and the dilation follows.
\end{proof}

\begin{cor}
Let 
$d=1+\lfloor \lg3\rfloor+\cdots+\lfloor\lg(n-1)\rfloor+ 
\lfloor\lg n \rfloor$ be the dimension of the binary hypercube 
$Q_d$. $Q_d$ can be embedded into the n-pancake with dilation $8$.
\end{cor}
\begin{proof}
$Q_d$ is a subgraph of the $2\times 3\times \cdots \times (n-1)\times n$ 
generalized hypercube.
\end{proof}
The latter representation of permutations yields an embedding 
of the GHC into the star graph.

\begin{thm}\label{star}
The $2\times 3\times \cdots \times (n-1)\times n$ generalized 
hypercube can be embedded into the star graph of dimension 
$n$ with dilation $4$.
\end{thm}
\begin{proof}
Again, the above representation of permutations is used. Let 
two permutations $X$ and $Y$, wherein at most three symbols 
have not the same location, i.e. $X=aAxByCzD$ and $Y=aAzBxCyD$. 
A path joining $X$ to $Y$ in the $n$-star is thus

$\qquad X=aAxByCzD\, \ra\, yAxBaCzD\, \ra\, zAxBaCyD\,$
 
$\qquad \qquad \ \ra\, xAzBaCyD\,\ra\, aAzBxCyD = Y$,

\smallskip \noindent
and the result follows.
\end{proof}

This last theorem improves on the result presented 
in~\cite{NSK90}. Indeed, Nigam {\sl et al.} show that the 
{\em binary} hypercube can be embedded into the $n$-star 
with dilation 4.Since the binary hypercube is a subgraph 
of the GHC, Theorem~\ref{star} generalizes that result.

\section{Conclusion}
We presented embeddings of rings, grids, and hypercubes 
into the pancake interconnection network. All embeddings 
have constant dilations, and some of them lead to similar 
results into the star graph. Possible improvements on the 
above results are twofold.
\begin{enumerate}
\item The embedding capabilities offered by the pancake 
interconnection network are very restrictive. In the present 
paper, the only embeddings of grids that are considered have 
size $n\times (n-1)!$ and $(n+(n-2)+(n-3)+\cdots+(p+1))\times p!$, 
for $p=2,\ldots,n-1$. Finding embeddings of $N_1\times N_2$ 
grids for all pairs $(N_1,N_2)$ such that $N_1 N_2\le n!$ 
would be a much more general result.
\item Some embeddings presented in the paper have congestion 
$O(n)$. A class of problems of the following kind remains open: 
find embeddings of the same interconnection networks with 
constant dilation and congestion, or else, show that such 
embeddings do not exist.
\end{enumerate}

\section{Acknowledgements}
Many thanks to Houssine Senoussi, who took great part 
in performing the first version of this work.

\end{document}